\documentclass[preprint,12pt]{elsarticle}
\usepackage{amssymb}
\usepackage{amssymb}
\usepackage{amsmath}
\usepackage{graphicx}
\usepackage{hyperref}
\usepackage{caption}
\usepackage{subcaption}
\usepackage{geometry}
\usepackage{multirow}
\usepackage{booktabs}
\usepackage{xcolor}
\begin{document}

\begin{frontmatter}

\title{Estimation of Electron Screening Potential in the $^{6}$Li(d,$\alpha$)$^{4}$He Reaction Using Multi-Layer Perceptron Neural Network}

\author[ipp]{D. Chattopadhyay\corref{cor1}}
\ead{dipayanchattopadhyay90@gmail.com}

\cortext[cor1]{Corresponding author}

\address[ipp]{Department of Physics, The ICFAI University Tripura, West Tripura 799210, India}

\begin{abstract}
Reactions between light charged nuclei at sub-Coulomb energies are crucial in astrophysical environments, but accurate cross-section measurements are hindered by electron screening. Traditional methods, including polynomial extrapolation and the Trojan Horse Method, often yield screening potentials exceeding adiabatic predictions. Building on the success of an MLP-based Artificial Neural Network (ANN) for the \( ^6\mathrm{Li}(p,\alpha)^3\mathrm{He} \) reaction~\cite{chattopadhyay24}, this work applies the same approach to the \( ^6\mathrm{Li}(d,\alpha)^4\mathrm{He} \) reaction. Experimental astrophysical S-factor data from literature are reanalyzed using the ANN to model the energy-dependent S-factor. The bare S-factor is extracted from data above 70~keV, where screening effects are minimal, and the screening potential is obtained by comparing with the low-energy region. The resulting screening potential is 147.95~eV, demonstrating the robustness of ANN-based methods for evaluating electron screening in low-energy nuclear reactions involving light nuclei.
\end{abstract}


\end{frontmatter}

\section{Introduction}
Investigations of nuclear reactions induced by light charged particles at \textit{Gamow energies} lie at the heart of modern nuclear astrophysics. At incident energies \( E \) well below the Coulomb barrier, the probability of nuclear fusion decreases exponentially, resulting in extremely small cross-sections. Precise measurements of these low-energy cross-sections are vital for refining our understanding of stellar evolution and the observed elemental abundances across the cosmos~\cite{rolfs88,adelberger11,engstler92,bertulani16,spitaleri16}.

To enable reliable extrapolation into the astrophysically relevant Gamow window, it is customary to express the fusion cross-section in terms of the \textit{astrophysical S-factor}:
\begin{equation}
\sigma(E) = \frac{1}{E} S(E) \exp\left(-2\pi\eta\right),
\end{equation}
where \( \eta = \frac{Z_1 Z_2 e^2}{\hbar v} \) is the Sommerfeld parameter, with \( Z_1 \) and \( Z_2 \) denoting the atomic numbers of the interacting nuclei, and \( v \) their relative velocity. For reactions involving light nuclei, \( S(E) \) varies slowly with energy, making it a more stable and practical quantity for extrapolation.

Equation~(1) assumes the fusion of bare nuclei. However, in laboratory experiments, nuclear targets are generally embedded in neutral atoms or molecules, which possess bound electrons. These electrons generate an electrostatic \textit{screening potential}, effectively reducing the Coulomb barrier between the projectile and the target nucleus. As a result, the projectile experiences an enhanced tunneling probability at low energies-a phenomenon referred to as \textit{laboratory electron screening}.

To incorporate this enhancement, the bare-nucleus cross-section \( \sigma_b(E) \) is modified as follows~\cite{engstler92}:
\begin{equation}
\sigma(E) = \sigma_b(E) \exp\left(\frac{U_e \pi \eta}{E}\right),
\end{equation}
where \( U_e \) is the electron screening potential. A rough estimate of \( U_e \) can be obtained using
\[
U_e = \frac{Z_1 Z_2 e^2}{R_a},
\]
where \( R_a \) is the atomic radius. This screening potential stems from the interaction between the incoming ion and the electron cloud surrounding the target nucleus.

The corresponding expression for the astrophysical S-factor becomes:
\begin{equation}
S(E) = S_b(E) \exp\left(\frac{U_e \pi \eta}{E}\right),
\end{equation}
where \( S_b(E) \) is the bare-nucleus S-factor. This relation highlights the need to correct laboratory-measured S-factors for electron screening effects to obtain values relevant to stellar conditions.

In stellar environments, nuclei reside in fully ionized plasmas where screening arises from collective Coulomb interactions rather than bound electrons. Each ion is surrounded by a Debye-H\"uckel screening cloud of oppositely charged particles, which reduces the effective Coulomb barrier. This \textit{plasma screening} depends on plasma temperature and density and differs fundamentally from laboratory electron screening.

The fusion cross-section in a stellar plasma can be expressed as:
\begin{equation}
\sigma_{pl}(E) = \sigma_b(E) f_{pl}(E),
\end{equation}
where the plasma enhancement factor is given by
\begin{equation}
f_{pl}(E) = \exp\left(\frac{U_{pl} \pi \eta}{E}\right),
\end{equation}
with \( U_{pl}(E) \) representing the plasma screening potential, often evaluated within the Debye-H\"uckel approximation.

These considerations are especially pertinent in the study of lithium nucleosynthesis. In particular, the reactions \( ^6\mathrm{Li}(p,\alpha)^3\mathrm{He} \) and \( ^6\mathrm{Li}(d,\alpha)^4\mathrm{He} \) dominate the destruction pathways of \( ^6\mathrm{Li} \), especially at energies \( E \lesssim 200\,\mathrm{keV} \), which are astrophysically relevant. Understanding their cross-sections is essential for resolving issues such as the anomalously low solar lithium abundance and discrepancies in Big Bang nucleosynthesis predictions~\cite{cruz08,spitaleri01}.

Engstler \textit{et al.}~\cite{engstler92} investigated the \( ^6\mathrm{Li}(d,\alpha)^4\mathrm{He} \) reaction and analyzed the S-factor data for both atomic and molecular targets using a third-order polynomial fit:
\[
S(E) = a + bE + cE^2 + dE^3.
\]

From their analysis, they extracted screening potentials of \( U_e = 380 \pm 250 \, \mathrm{eV} \) for atomic targets and \( U_e = 330 \pm 120 \, \mathrm{eV} \) for molecular targets-both significantly higher than the theoretically predicted adiabatic limit of \( U_e = 175 \, \mathrm{eV} \). Additional studies using the Trojan Horse Method (THM)~\cite{spitaleri01} inferred a screening potential of \( U_e = 340 \pm 51 \, \mathrm{eV} \), while Barker~\cite{barker02} reported \(U_e = 248\)\,eV based on polynomial fits to published data. Ruprecht \textit{et al.}~\cite{ruprecht04} reported a value of \( U_e = 190 \pm 50 \, \mathrm{eV} \), attributing the reduction to a \( 2^+ \) subthreshold resonance. More recently, an R-matrix analysis by Grineviciute \textit{et al.}~\cite{grineviciute15} yielded \( U_e = 149.5 \, \mathrm{eV} \), further illustrating the variability and uncertainty in existing determinations of the screening potential.

In parallel with these efforts, Artificial Neural Networks (ANNs) have emerged as powerful tools for modeling complex and nonlinear relationships between variables. Their adaptability and predictive capabilities have enabled their application across a wide range of scientific domains, including nuclear physics. In recent years, ANNs and other Machine Learning (ML) methods have been employed for a variety of nuclear physics tasks: constructing detector response models~\cite{akkoyun13}; evaluating fusion suppression in weakly bound systems~\cite{chattopadhyay25}; estimating fusion cross-sections~\cite{akkoyun20}; charge identification from ionization data~\cite{XiangLei23}; predicting ion ranges~\cite{Minagawa24}; determining elemental abundances~\cite{Cohen24}; modeling stopping powers~\cite{Parfitt20}; forecasting $\alpha$-decay half-lives~\cite{Ma23}; global charge density modeling~\cite{Shang24}; investigating mirror charge radii~\cite{Zhang24}; emulating spontaneous fission~\cite{Lay24}; calculating nuclear binding energies~\cite{Zeng24}; constraining Woods--Saxon potentials~\cite{Gao24}; predicting separation energies~\cite{athanassopoulos04}; nuclear mass systematics~\cite{athanassopoulos004}; learning mass correlations~\cite{sharma22}; computing ground-state energies~\cite{bayram14}; impact parameter estimation in heavy-ion collisions~\cite{david95,bass96,haddad97}; $\beta$-decay energy predictions~\cite{akkoyun14}; and rms charge radius estimation~\cite{akkoyun013,Wu20}.

Building on the success of using ANNs to estimate the screening potential in the \( ^6\mathrm{Li}(p,\alpha)^3\mathrm{He} \) reaction~\cite{chattopadhyay24}, the present work extends this methodology to the \( ^6\mathrm{Li}(d,\alpha)^4\mathrm{He} \) reaction. A multi-layer perceptron (MLP) architecture is trained using experimental data compiled from the literature~\cite{engstler92,elwyn77}, enabling prediction of the total astrophysical S-factor across the low-energy domain. The electron screening potential is then extracted by evaluating the ratio of the predicted total S-factor to the bare S-factor.

Details of the ANN methodology are discussed in the following section, along with the extraction of the screening potential and subsequent conclusions.

\section{Artificial Neural Network}

An Artificial Neural Network (ANN) is a computational framework inspired by the architecture and operation of biological neural networks, especially the human brain. It is designed to emulate the brain’s capacity to process information, recognize patterns, learn from data, and make predictive decisions~\cite{Ma23,Shang24,Zhang24,Lay24,Zeng24,Gao24,Wu20,Wolfgruber24,Li24,Yang23}. The core unit of an ANN is the \textit{neuron}, which performs a weighted summation of its inputs, applies a nonlinear activation function, and passes the result forward.

In this work, we employ a Physics-Informed Neural Network (PINN)-enhanced approach, where prior physical knowledge is incorporated into the learning pipeline through polynomial feature transformation. Specifically, polynomial basis functions up to the second degree are used to map the input energy values to a higher-dimensional feature space, thereby embedding smooth, physics-inspired representations into the model input. This expansion aids in capturing nonlinear trends that are consistent with physical expectations of the S-factor's energy dependence.

The dataset used for training and testing comprises experimental measurements of the \( ^6\mathrm{Li}(d,\alpha)^4\mathrm{He} \) reaction collected from established sources~\cite{engstler92,spitaleri01}. Each sample consists of the center-of-mass energy (\(E\)) as the input feature and the corresponding astrophysical S-factor (\(S\)) as the output variable. The dataset is randomly split into 70\% training and 30\% testing subsets to balance model learning and validation on unseen data.

\begin{figure}[h]
\centering
\includegraphics[scale=0.45]{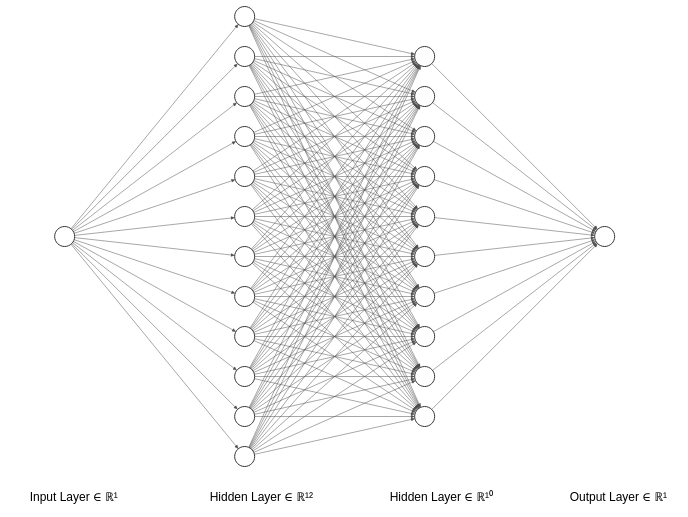}
\caption{\label{fig:ann_next} Schematic representation of a typical Feed Forward Neural Network architecture, depicting the structure and arrangement of input, hidden, and output layers along with the connections between them.}
\end{figure}

The network follows a feedforward structure, the most widely used class of ANN, where information flows unidirectionally from the input layer to the output layer through a sequence of hidden layers. The ANN is structured into an \textit{input layer}, one or more \textit{hidden layers} for hierarchical feature extraction, and a \textit{single output layer} that predicts the S-factor. A schematic of the architecture is shown in Fig.~\ref{fig:ann_next}, created using the NNSVG tool~\cite{nnsvg}.

The ANN architecture adopted in this study is consistent with our prior work~\cite{chattopadhyay24}, where a Multi-Layer Perceptron (MLP) configuration was successfully used to predict the electron screening potential in the \( ^6\mathrm{Li}(p,\alpha)^3\mathrm{He} \) reaction. Here, the same framework is extended to the \( ^6\mathrm{Li}(d,\alpha)^4\mathrm{He} \) system. The model is initialized with one input layer, two hidden layers and one output layer. The output layer is configured to produce a single continuous value representing the predicted S-factor.

\begin{table} [h]
\centering
\caption{The electron screening potential \( U_e(\text{Lab}) \) for the \( ^6\text{Li}(d,\alpha)^4\text{He} \) reaction, reported by various studies~\cite{engstler92,spitaleri01,barker02,ruprecht04,grineviciute15}, is compared with the value extracted in the present work using a Multi-Layer Perceptron (MLP)-based Artificial Neural Network (ANN) model.}
\label{tab3}
\vspace*{0.1cm}
\begin{tabular}{cccc}\hline \hline
&&&\\
U$_{e}$(Lab) & $\Delta$U$_{e}$(Lab) & Method & References\\
(eV) & (eV) &  &\\
&&& \\ \hline
&&& \\
380&$\pm$250& Polynomial fitting of Atomic Target &~\cite{engstler92}\\
&&& \\
330&$\pm$120& Polynomial fitting of Molecular Target &~\cite{engstler92}\\
&&& \\
248&$---$&Polyn. Fit&~\cite{barker02}\\
&&& \\
340&$\pm$51&THM&~\cite{spitaleri01}\\
&&& \\
190&$\pm$50& &~\cite{ruprecht04}\\
&&& \\
149.5&$---$&R-MATRIX &~\cite{grineviciute15}\\
&&& \\
147.95&$---$& ANN &Present Work\\
&&& \\ \hline
\end{tabular}
\end{table}

\begin{figure}[h]
\centering
\includegraphics[scale=0.6]{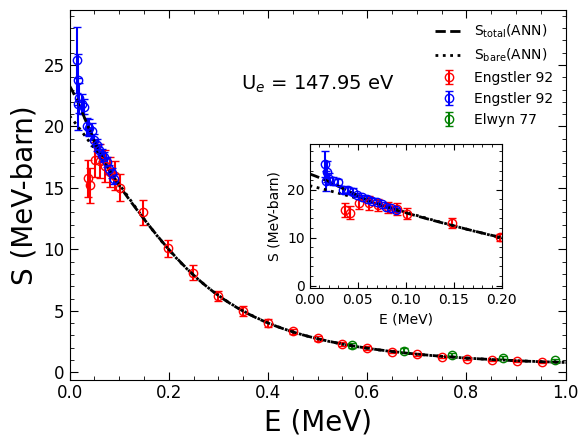}
\caption{\label{fig:screening} Comparison between the experimental data~\cite{engstler92,spitaleri01} and the total (screened) S-factor and bare S-factor predicted by the present ANN-based model. The inset highlights the low-energy region from 0.0 to 0.2~MeV, where the short-dashed curve represents the screened (total) S-factor and the dotted curve denotes the bare S-factor, clearly illustrating the enhancement due to electron screening.}
\end{figure}
\begin{figure}[h]
\centering
\includegraphics[scale=0.6]{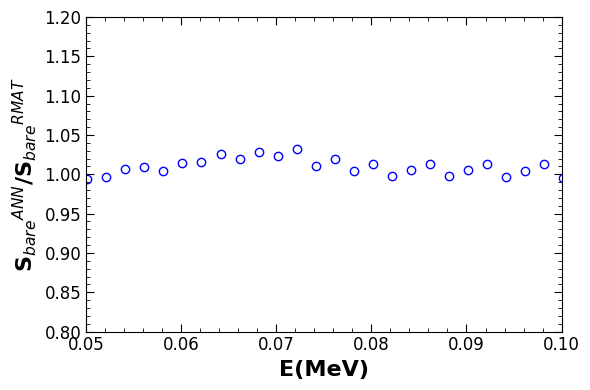}
\caption{\label{fig:sbare_rmat} Comparison of the bare S-factor obtained from the present ANN model with the R-matrix calculation~\cite{grineviciute15}.}
\end{figure}

\begin{figure}[h]
\centering
\includegraphics[scale=0.6]{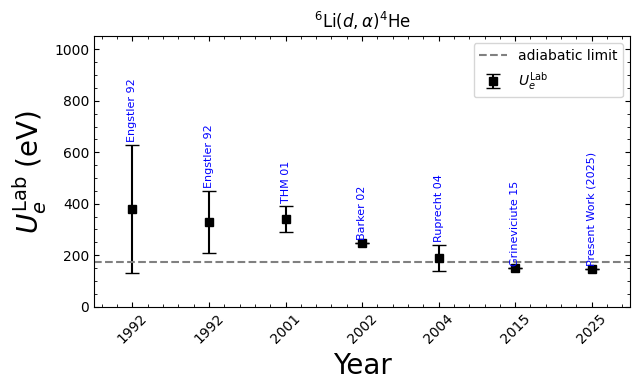}
\caption{\label{fig:Ue_year} Comparison of the electron screening potential obtained from the present calculation with previously published values from the literature~\cite{engstler92,spitaleri01,barker02,ruprecht04,grineviciute15}.}
\end{figure}

To determine the optimal ANN architecture, a Randomized Search was performed over a broad hyperparameter space. This included various hidden layer configurations such as \((32, 16)\) and \((64, 32)\); activation functions (\texttt{ReLU}, \texttt{tanh}, \texttt{logistic}); regularization strengths (\(\alpha = 0.1, 1.0, 10.0\)); learning rate strategies (\texttt{constant}, \texttt{adaptive}); and optimization algorithms (\texttt{Adam}, \texttt{SGD}, and \texttt{LBFGS}). This comprehensive search allowed for systematic evaluation of model performance under diverse training dynamics.

Each hyperparameter configuration was evaluated using five-fold cross-validation to ensure reliable model selection and to mitigate overfitting. The final optimized model employed the \texttt{LBFGS} solver, which was found to yield the best performance. The selected configuration included hidden layers of sizes \((32, 16)\), the \texttt{tanh} activation function, an adaptive learning rate with an initial value of 0.001, and a regularization strength of \(\alpha = 0.1\). Additional optimal settings included early stopping enabled, a validation fraction of 0.2, and numerical stability parameters \(\beta_1 = 0.8\), \(\beta_2 = 0.98\), and \(\epsilon = 10^{-8}\). The model was trained for a maximum of 6,000 iterations.

Model performance was evaluated using the Mean Squared Error (MSE) and the coefficient of determination (\(R^2\)). The trained network demonstrated excellent predictive performance, achieving low normalized Mean Squared Errors (MSEs) and high \( R^2 \) scores on both the training and testing datasets. These results highlight the effectiveness of the physics-informed feature engineering approach in capturing the complex, nonlinear behavior of astrophysical S-factors in light-ion fusion reactions.

\section{Calculations of Electron Screening Potential}
A Multi-Layer Perceptron (MLP)-based Artificial Neural Network (ANN) algorithm is employed to extract the bare and screened astrophysical S-factor from experimental data reported in the literature. The experimental datasets used in this study are obtained from the {\sc EXFOR} database~\cite{engstler92,spitaleri01}, covering a range of center-of-mass energies that includes both the screened region (below 70~keV) and the unscreened region (above 70~keV), where the effects of electron screening are negligible. The compiled dataset was randomly partitioned into training and testing subsets using a 70:30 split ratio. The ANN model was trained on the training set, and through systematic experimentation, the optimal architecture was determined. Hyperparameters were fine-tuned using cross-validation and randomized search, targeting an optimal configuration that maximizes the correlation between predicted and actual S-factor values. The resulting high correlation coefficient serves as an indicator of both the model’s predictive accuracy and its ability to minimize uncertainty in extrapolating the bare S-factor.

The final architecture of the Artificial Neural Network (ANN) model consists of one input layer, two hidden layers with 32 and 16 neurons respectively, and a single output layer. This configuration was identified as optimal through a randomized hyperparameter search focused on minimizing prediction error and improving generalization. The model was trained using the Limited-memory Broyden-Fletcher-Goldfarb-Shanno (\texttt{LBFGS}) solver, with a maximum of 6,000 iterations. The training process employed an adaptive learning rate initialized at 0.001, a regularization parameter \(\alpha = 0.1\), and the \texttt{tanh} activation function. Early stopping was enabled with a validation fraction of 0.2 to prevent overfitting. Following training, the same architecture and hyperparameter configuration were used to evaluate model performance on the test dataset. In the second phase of analysis, experimental data corresponding to center-of-mass energies above 70~keV—where the effects of electron screening are negligible—were used to train the ANN to extract the bare astrophysical S-factor. The electron screening potential was subsequently determined by taking the ratio of the experimentally observed (screened) S-factor to the bare S-factor predicted by the trained ANN model.

The screened and bare astrophysical S-factors predicted by the present Artificial Neural Network (ANN) model are depicted in Fig.~\ref{fig:screening}. The extrapolated bare S-factor at zero energy, as predicted by the trained ANN model, is 20.86~MeV-barn. This result is compared with the value obtained from the R-matrix calculation reported in Ref.~\cite{grineviciute15}, as illustrated in Fig.~\ref{fig:sbare_rmat}. The ANN-based estimate shows good agreement with the R-matrix result, demonstrating the reliability of the present approach. The electron screening potential derived from the ANN analysis is found to be \( U_e = 147.95 \)~eV, as given in Table~\ref{tab3}, and is compared with previous estimations in Fig.~\ref{fig:Ue_year}. Notably, the obtained value is in good agreement with results from R-matrix analyses, remaining well within quoted uncertainty limits. The performance of the model was quantitatively assessed using both the training and testing subsets. It achieved a coefficient of determination \( R^2 \) of 0.982 on the training set and 0.971 on the testing set, indicating strong predictive accuracy. The Normalized Mean Squared Error (NMSE) was calculated to be 4.3\% for training and 5\% for testing datasets. These results demonstrate that the Multi-Layer Perceptron-based ANN model offers a viable and efficient alternative approach for extracting the electron screening potential in light charged-particle induced nuclear reactions of astrophysical relevance.

\section{Conclusions}

In this study, a Multi-Layer Perceptron (MLP)-based Artificial Neural Network (ANN) model was employed to extract the electron screening potential in the \( ^6\text{Li}(d,\alpha)^4\text{He} \) reaction. By training the model on experimental astrophysical S-factor data from the {\sc EXFOR} database, the ANN successfully predicted both the bare and screened S-factors across a relevant low-energy range. The bare S-factor was estimated using data above 70~keV, where electron screening effects are negligible, and the screening potential was subsequently determined by comparing the predicted bare S-factor to the total S-factor.

The ANN model, optimized via randomized hyperparameter search and five-fold cross-validation, demonstrated strong predictive accuracy, achieving \( R^2 \) values of 0.982 and 0.971 for the training and testing datasets, respectively. The bare S-factor at zero energy, extrapolated from the model, was found to be \( S_b(0) = 20.86 \, \text{MeV-barn} \), while the corresponding electron screening potential was determined as \( U_e = 147.95 \, \text{eV} \). This value agrees well with earlier R-matrix calculations and lies within the range of previously reported experimental and theoretical estimates.

These findings highlight the utility of ANN-based approaches as robust and data-driven alternatives to traditional methods for estimating electron screening effects in low-energy nuclear reactions. The methodology is particularly valuable for reactions involving light nuclei at sub-Coulomb energies, where conventional techniques often face challenges. Overall, this work demonstrates the potential of machine learning, and neural networks in particular, as reliable tools in nuclear astrophysics research.

\end{document}